\documentclass[aps,prl,twocolumn,showpacs,groupedaddress,letterpaper]{revtex4}  
\usepackage{graphicx}  

\newcommand{\beq}{\begin{equation}}
\newcommand{\eeq}{\end{equation}}
\newcommand{\bea}{\begin{eqnarray}}
\newcommand{\eea}{\end{eqnarray}}

\begin{document}


\title{Radial distribution of RNA genome packaged inside spherical viruses}


\author{Se Il Lee and T. T. Nguyen}
\affiliation{Georgia Institute of Technology, School of Physics,
837 State Street, Atlanta, Georgia 30332-0430}


\date{\today}

\begin{abstract}
The problem of RNA genomes packaged inside spherical viruses is studied.
The viral capsid is modeled as a hollowed sphere. The attraction between 
RNA molecules and the inner viral capsid is assumed to be non-specific
and occurs at the inner capsid surface only. For small capsid attraction,
it is found that monomer concentration of RNA molecules
is maximum at the center of the capsid to maximize their
configurational entropy. For stronger capsid attraction, 
RNA concentration peaks at some distance near the capsid.
In the latter case, the competition between the branching of RNA secondary
struture and its adsorption to the inner capsid results in the formation
of a dense layer of RNA near capsid surface. The layer thickness is a slowly
varying (logarithmic) function of the capsid inner radius. Consequently, 
for immediate strength of RNA-capsid interaction,
the amount of RNA packaged inside a virus is proportional to the capsid 
{\em area} (or the number of proteins) instead of its volume. 
The numerical profiles describe reasonably well the experimentally observed 
RNA nucleotide concentration profiles of various viruses.
\end{abstract}

\pacs{81.16.Dn, 87.16.A-, 87.19.rm}

\maketitle

Viruses attract broad interests from physics community due to their
ability of spontaneous self assembly. Many viruses can be produced 
both {\em in-vivo} and {\em in-vitro} as highly robust and monodisperse
particles. As a result, beside biomedical applications, 
understanding virus assembly can also have novel promising
applications in nanofabrication. At the basic level,
viruses consist of viral genomes (RNA or DNA molecules) packaged
inside a protective protein shell (viral capsid). The structures of viral
capsids for most viruses are well understood from high-resolution experiments
using cryoelectron microscopy or X-ray analysis \cite{Johnson98, JohnsonBook},
as well as theoretical studies \cite{Lidmar, NguyenPRL062}. Single-stranded
RNA (ssRNA) viruses also package their genome spontanously during
assembly. Several theoretical studies have demonstrated that the interaction between
capsid proteins and RNA nucleotide basis plays an important role 
in the RNA packaging process, both energetically and kinetically
\cite{BorisRNAkinetic,BorisarXiv,BelyiPNAS,BruinsmaRNAdodecahedron,BruinsmaSchoot05}.
However, unlike the structural study
of viral capsid, there is still a lack of general understanding of 
struture of packaged RNA. In references \onlinecite{BelyiPNAS, BorisarXiv,BruinsmaSchoot05},
different models of RNA packaging inside viruses were studied. However, 
all these works treat RNA molecules as {\em linear} flexible polymers. 
In this letter, 
we want to address the question of how RNA molecules are arranged
inside a spherical virus, {\em explicitly} taking into acount
the {\em branching} degree of freedom of RNA secondary structure. 

\begin{figure}[th]
\begin{center}
\includegraphics[width=8.5cm]{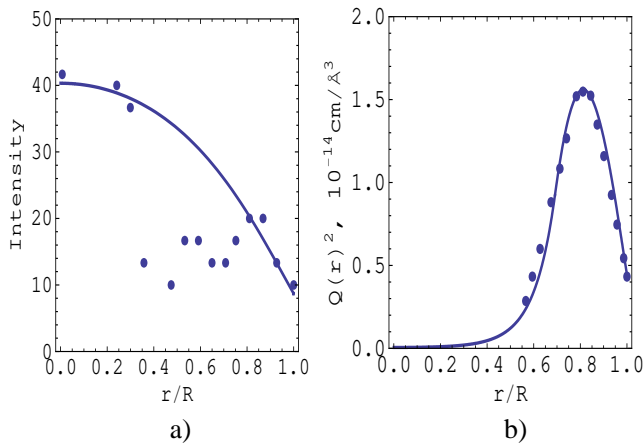}
\end{center}
\caption{Two different profiles for RNA monomer concentration inside spherical
viruses. Points are experimental data and solid lines are theoretical
fit. a) Profile II, Eq. (\ref{Eq:profile2}),
fitted to RNA concentration of Dengue virus obtained from
cryoelectron microscopy experiment \cite{RossmannDengue02}. b)
Profile III, Eq. (\ref{Eq:profile3}),
fitted to RNA concentration of bacteriophage MS$_2$ obtained from
small angle neutron scattering experiment \cite{Jacrot77}.
}
\label{fig:aProfile}
\end{figure}

We focus on a particular class of ssRNA viruses where the 
interaction between capsid proteins and RNA molecules is non$-$specific and occurs
dominantly at the inner surface of the capsid. This is the case
for viruses where basic amino acids are located on the surface and
electrostatic interaction is strongly screened in the bulk solution
(examples of such viruses are bacteriophage MS$_2$, Q Beta,
Dengue, Immature Yellow Fever,... generally viruses belonging to
group B and C mentioned in Ref. \onlinecite{BorisarXiv}).
(In some viruses such as pariacoto virus\cite{JohnsonDodecahedralRNA}, 
the viral capsid forces some fraction of RNA molecules
to adopt it dodecahedron structure.
In that case, the theory presented below should be applied to the 
free fraction of these RNAs.)
 Even though RNA-capsid interaction
only occurs at the surface, RNA radial
concentration profiles and the amount of RNA packaged inside
a virus can be dictated by the strength of this interaction. 
The main result of this papers is that there are
two different profiles for the radial RNA nucleotide concentration. 
For small capsid attraction,
the RNA concentration is maximum at the center of the capsid. A representative
virus (the Dengue virus) for this profile is shown in Fig. \ref{fig:aProfile}a.
For larger capsid attraction, the RNA concentration is maximum at 
a distance close to (but always smaller than) the inner capsid radius.
A representative virus (the bacteriophage MS$_2$) for this profile is
shown in Fig. \ref{fig:aProfile}b.
For the later case, the RNA molecules form a dense layer at the inner capsid
surface. The thickness of this layer varies very slowly (logarithmic) with the 
capsid radius. As a result, the amount of RNA packaged inside such viruses is
proportional to the capsid area (or the number of capsid proteins) instead of
its volume.



\begin{figure}[th]
\resizebox{5cm}{!}{\includegraphics{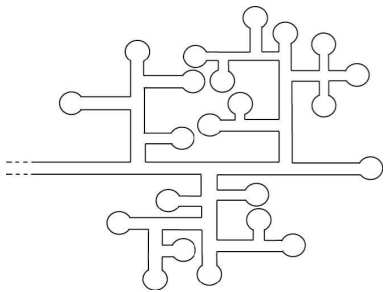}}
\caption{Schematic representation of the secondary structure of a
single-stranded RNA molecule as a collection of linear sections,
branch-points, and end-points. The molecule can freely fluctuate
between different branching configurations.}
\label{fig:RNAschematic}
\end{figure}

It is well known that ssRNA molecules fold on themselves due to 
base-pairing interaction between their nucleotides. Because nucleotide
sequence of ssRNA molecules is not perfect for such pairing, their
secondary structure is highly nonlinear. 
To the first approximation, RNA molecules are considered to be highly flexible branch polymers
which can fluctuate freely over all possible branching configurations. 
Different branching configurations are described in the schematic way shown in 
Fig. \ref{fig:RNAschematic}, characterized by fugacities
for ``bi-functional" units (linear sequences), ``tri-functional units"
(branching points) and ``endpoints" (stem-loops or hair-pins). 
We assume good solvent condition with repulsive interactions between
the different units (with no ``tertiary" pairing). 
Using a mean-field approximation\cite{NguyenPRL06}
to a field theory for solutions of branching polymers of this 
type\cite{LIPRA}, one can write down an expression
for the free energy density of RNA solution $W[Q(\vec{r})]$ as
\beq
\frac{W[Q(\vec{r})]}{m}=\frac{\epsilon}{2} Q(\vec{r})^2-
	\frac{w}{6}Q(\vec{r})^3
	+muQ(\vec{r})^4-hQ(\vec{r})~,
\label{Eq:WQ}
\eeq
where $\epsilon$, $w$, $h$ and $m$ are the fugacity of the monomers,
branch points, the end-points and the whole polymers respectively.
The coefficient $u$ is proportional to the second-order
virial coefficient for monomer-monomer interaction (since RNA molecules
are assumed to be in good solvent, $u$ is positive). $Q(\vec{r})$ is the
order parameter of the field theory and is proportional to the
concentration of end-points. Note that if one sets $w=0$ 
(the branching degree of freedom is suppressed), Eq. (\ref{Eq:WQ})
recovers the well known expression for the free energy density of a 
solution of linear polymers\cite{deGennesBook}. Based on this mean-field
expression, it is suggested that RNA are prone to a surface condensation which
is different from that of linear polymer\cite{NguyenPRL06}.
In this paper, we will use the mean-field expression, Eq. (\ref{Eq:WQ}),
 to study how RNA
molecules are packaged inside a virus. For simplicity, we model the
viral capsid as a hollow sphere with inner radius $R$. We also assume
that RNA molecules are radially distributed inside the capsid
so that $Q(\vec{r}) \equiv Q(r)$ where $r$ is the radial distance
from the center of viral capsid. As a result, the excess free energy
of the RNA molecules packaged inside a capsid can be written as
\beq
H_{\mbox{MF}}=H_s+\int_0^R 4\pi r^2dr\left\{
	\frac{m}{2}\left(\frac{dQ}{dr}\right)^2+\Delta W\right\},
\label{Eq:HMF}
\eeq
with $\Delta W[Q(r)] = W[Q(r)] - W[Q_{bulk}]$. 
The first term in Eq. (\ref{Eq:HMF}) denotes the interaction energy 
of the capsid proteins with the RNA molecules. Assuming this interaction
occurs only at the inner capsid surface, $H_s$ can be written as the sum of
contributions from monomers and endpoint adsorptions:
\beq
H_s = 4\pi R^2 m [-\gamma_1 Q(R) - \gamma_2 Q(R)^2/2 ] ,
\eeq
where $\gamma_{1,2}$ are the strengths of the adsorption.

Due to the cubic term proportional to $w$ in Eq. (\ref{Eq:WQ}),
for small positive $\epsilon$, the free energy density
$W(Q)$ has two minima, $Q_D$ and $Q_C$, corresponding to,
respectively, the mean-field order parameter of a dilute bulk RNA 
solution and that of a condensed bulk RNA solution. 
A first-order condensation transition takes 
place when $W(Q_D) = W(Q_C)$. We will always assume RNA solution lies at this
coexistence regime so that both the dilute
and dense phases of RNA solution are close in energy. Therefore,
we set bulk value $Q_{bulk} = Q_D$. The equilibrium RNA concentration
profile corresponds to the profile $Q(r)$ that 
minimizes the Hamiltionian Eq. (\ref{Eq:HMF}). Setting the
functional derivative, $\delta H_{MF}/\delta Q$ to zero,
 we obtain the Euler-Lagrangian equation
\begin{eqnarray}
\frac{d^2Q}{dr^2}+\frac{2}{r}\frac{dQ}{dr}-\frac{1}{m}\frac{\partial\Delta W}{\partial Q}=0
,
\label {Eq:Euler}
\end{eqnarray}
and a boundary condition at the inner capsid surface:
\begin{equation}
    \left.\frac{dQ}{dr}\right|_{r=R} =\frac{{H_s}^{'}[Q(R)]}{4\pi R^2m}
	=-\gamma_1-\gamma_2 Q(R) ~.
\label {Eq:Boundary}
\end{equation}

To proceed further, we approximate $\Delta W$ using 
the double parabolic potential form\cite{ColloidWetting}:
\begin{equation}
\Delta W(Q) = \left \{ 
   \begin{array}{rl}
	\frac{1}{2}m\lambda_D^2 (Q-Q_D)^2 \text{ for } Q < Q_m\\
	\frac{1}{2}m\lambda_C^2 (Q-Q_C)^2 \text{ for } Q > Q_m
   \end{array} \right.
,
\label {Eq:soe}
\end{equation}
where 
$
Q_m = (\lambda_D Q_D + \lambda_C Q_C)/(\lambda_D + \lambda_C)
$
is the point
where the two parabolas cross each other forming a cusp.
The two coefficients $\lambda_D^2,\lambda_C^2$ are the stiffness of
the free energy density of RNA solution near the two minima. They
are proportional to the corresponding correlation lengths of
the two phases. In general, this double parabolic potential form 
for the free energy density breaks down near the critical
temperature where the first order transition becomes second order,
or when the fugacity of branch points, $w$, goes to 0
(the branching degree of freedom is suppressed and
RNA molecules behave as a linear polymer). However, it was shown
\cite{NguyenPRL06} that the mean-field expression, Eq. (\ref{Eq:WQ}), 
breaks down before this limit is approached. If one stays 
within the limit of mean-field theory, the double parabola
approximation is a reasonable approximation.
We will come back to its limitation in later discussion.
With this approximate form of $\Delta W$, Eq. (\ref{Eq:Euler}) becomes
linear and easy to solve. The general solution is a linear combination of
$\exp(\pm \lambda_{D,C} r)/r$. There are three possible
concentration profiles for the RNA molecules.

{\em Profile I}. If for all $r$, $Q(r) < Q_m$, then the solution to the Euler
equation is 
\beq
Q(r)=-C_{10}{\sinh(\lambda_D r)}/({\lambda_D r})+Q_D~,
\label{Eq:profile1}
\eeq
where 
\beq
C_{10}=\frac{(\gamma_1+\gamma_2 Q_D) R }
{\cosh(\lambda_D R)+(\gamma_2 R-1)\sinh(\lambda_D R)/(\lambda_D R)} .
\eeq
Because the interaction of the RNA monomers with the 
viral capsid is attractive, $\gamma_{1,2} >0$, the coefficient
$C_{10}$ is a positive quantity. According to Eq. (\ref{Eq:profile1}),
this means that for all $r$,
the endpoint (and monomer) concentration in this profile
is always smaller than the bulk value, $Q(r) < Q_D = Q_{bulk}$. 
This is a non physical situation. Therefore, we discard this solution
from later consideration. 

{\em Profile II}. The second possibility is the case that for all $r$, $Q > Q_m$.
Accordingly, the solution is
\begin{eqnarray}
Q(r)=-C_{20}{\sinh(\lambda_C r)}/({\lambda_C r})+Q_C ~,
\label{Eq:profile2}
\end{eqnarray}
where 
\beq
C_{20}=\frac{(\gamma_1+\gamma_2 Q_C) R }
{\cosh(\lambda_C R)+(\gamma_2 R-1)\sinh(\lambda_C R)/(\lambda_C R)}
~.
\eeq
This solution is a monotonously decreasing function of $r$
and the RNA concentration is maximum at the {\em center}
of the capsid. Because of the requirement that $Q(R)$ must 
be greater than $Q_m$,
this profile is possible only for very weak adsorption
(in practice, $\lambda_C R \gg 1$, this requirement means
$(\gamma_1/Q_C+\gamma_2)/\lambda_D < 1$). As a result,
RNA monomers want to concentrate at the center of the capsid to
gain their configurational entropy (minimizing the gradient
term in Eq. (\ref{Eq:HMF}) ).

{\em Profile III}. The third possibility is that $Q(r)$ passes through $Q_m$ at
some distant $r=r_0$ ($0<r_0<R$) such that $Q(r=r_0)=Q_m$.
We can interpret $r_0$ as the boundary between the dilute and
the condensed phases of RNA molecules inside the capsid. 
Requiring the density profile $Q(r)$ and its derivative 
$Q^{'}(r)$ to be continuous at $r_0$, we get
\begin{equation}
Q(r) = \left \{ 
   \begin{array}{ll}
   (Q_0 -Q_D) \frac{\sinh(\lambda_D r)}{\lambda_D r} +Q_D~~~ \text{ for }  r<r_0  \\
C_{31}\frac{\exp(\lambda_C r)}{\lambda_Cr} + C_{32} \frac{\exp(-\lambda_C r)}{\lambda_Cr}
 + Q_C ~ \text{ for } r_0 < r < R
   \end{array} \right.
\label {Eq:profile3}
\end{equation}
where $Q_0=Q(0)$ and
\begin{eqnarray*}
C_{31}&=&-{\exp[-(\lambda_C +\lambda_D)r_0]}(Q_0 -Q_D)(\lambda_C/\lambda_D-1)/4
\\
&&+\exp[-(\lambda_C -\lambda_D)r_0](Q_0 -Q_D)(\lambda_C/\lambda_D+1)/4 \\
&&-\exp(-\lambda_C r_0)(Q_C -Q_D)(\lambda_Cr_0 +1)/2 ,\\
C_{32}&=&\exp[(\lambda_C +\lambda_D)r_0](Q_0 -Q_D)(\lambda_C/\lambda_D-1)/4
\\&&
-\exp[(\lambda_C -\lambda_D)r_0](Q_0 -Q_D)(\lambda_C/\lambda_D+1)/4 \\
&&-\exp(\lambda_C r_0)(Q_C -Q_D)(\lambda_Cr_0 -1)/2 .
\end{eqnarray*}
$r_0$ and $Q_0$ are two unknowns in the solution above. They can be solved
by matching the boundary condition, Eq. (\ref{Eq:Boundary}), and
the condition $Q(r_0) = Q_m$. The later condition gives
\beq
Q_0 = Q_D+ (Q_C-Q_D)
\frac{\lambda_C}{\lambda_C+\lambda_D}
 \frac{\lambda_D r_0}{\sinh(\lambda_D r_0)}~.
\label {Eq:Q0}
\eeq
Substituting Eq. (\ref{Eq:profile3}) and (\ref{Eq:Q0}) into
the boundary condition Eq. (\ref{Eq:Boundary}), we arrive
at the equation for $r_0$:
\bea
&&
\left(1+\frac{\gamma_2}{\lambda_C}-\frac{1}{\lambda_C R}\right)\left[-
\frac{\lambda_D r_0 \exp(-\lambda_D r_0)}{\sinh(\lambda_D r_0)}
+1+\frac{\lambda_D}{\lambda_C}
\right] u^2 
\nonumber\\
&&- 2\lambda_s R u
\nonumber \\
&&
- \left(1-\frac{\gamma_2}{\lambda_C}+\frac{1}{\lambda_C R}\right)\left[
 \frac{\lambda_D r_0\exp(\lambda_D r_0)}{\sinh(\lambda_D r_0)}
-1-\frac{\lambda_D}{\lambda_C}
\right]  = 0 ,
\nonumber \\
\label{Eq:r0}
\eea
where
%
$
u = \exp[\lambda_C (R-r_0)]~.
$
%
The parameter
\beq
\lambda_s = (1+\lambda_D/\lambda_C)(\gamma_1+\gamma_2Q_C)/(Q_C - Q_D)~,
\eeq
is proportional to the strength of RNA adsorption at the
inner capsid surface and has dimension of inverse length.
Obtaining an analytical solution for $r_0$ from
Eq. (\ref{Eq:r0}) is a highly non-trivial task and numerical
solution is generally needed. Nevertheless, 
we can understand important qualitative 
features of the RNA concentration profile 
by solving for $r_0$ in the limit of strong capsid
RNA adsorption ($\lambda_s R \gg 1$) and small 
correlation length of RNA concentrated phase 
($\lambda_C R \gg 1$). In this limit, the first two
terms in Eq. (\ref{Eq:r0}) are the two most dominant
ones. Balancing them, we get $u \simeq 2 \lambda_s R$, or
\beq
r_0 \simeq R - {\lambda_C}^{-1} \ln (2\lambda_s R) .
\label{Eq:r0solution}
\eeq
As we mentioned above, $r_0$ can be considered as the boundary 
between a dense RNA phase near the capsid
and a dilute RNA phase at the capsid center. The quantity $d=R-r_0$, therefore, can be considered
the thickness of this dense RNA layer. According Eq. (\ref{Eq:r0solution}), $d \propto \ln R$
which is {\em parametrically smaller} than the capsid radius, $R$
\footnote{
The $\ln R$ dependence of
$d$ is also obtained for the wetting layer on the surface of
a colloid \cite{ColloidWetting}. This is to be expected because
Eq. (\ref{Eq:WQ}) can be mapped onto the Cahn theory
of wetting transition\cite{NguyenPRL06, deGennesWetting}.
Here we show that the negative curvature of the inner viral capsid apparently does
not significantly affect this logarithmic dependency.}. 
In other words, our RNA concentration
profile shows a dense RNA layer condensed on the inner
capsid with thickness which varies {\em very slowly} with its radius.
Consequently, the amount of RNA packaged inside
the virus is proportional to the capsid {\em area}
(or the number of capsid proteins)
instead of its volume. 
In recent works \cite{BelyiPNAS,BorisarXiv},
a similar dependence is observed when positively charged
amino acids of capsid proteins are located in their long
flexible peptide arms. In their works, the thickness of
RNA molecules (treated as {\em linear} polymers)
layer depends on the length of these arms. On the other
hand, for the class of viruses we study in this paper
where the basic amino acids are located at the inner capsid
surface instead of peptide arms, the competition between
the {\em branching} degree of freedom of the secondary structure of RNA
molecules and the attraction of capsid proteins is responsible
for the layer structure and the thickness scales as $\ln R$.
Another interesting feature of RNA concentration profile III is
the fact that it does not peak at the inner capsid radius
$R$ but at some smaller radius. This is the direct
consequence of the boundary condition, Eq. (\ref{Eq:Boundary})
which forces the RNA concentration to decrease
in the vicinity of the capsid. 

In Fig. \ref{fig:aProfile}, we plot examples of the two profiles,
Eq. (\ref{Eq:profile2}) and Eq. (\ref{Eq:profile3}),
fitted to the experimental data for two viruses, 
the Dengue virus and bacteriophage MS$_2$. 
The data for the Dengue virus was obtained using cryoelectron microscopy
\cite{RossmannDengue02}. The data for bacteriophage MS$_2$ was
obtained using small angle neutron scattering
measurements \cite{Jacrot77}. Both viruses have
most of their basic amino acids located on the surface of inner capsid,
therefore our model capsid can be used.  Both theoretical profiles
show reasonable agreement with experiment results.

So far, when solving the Euler-Lagrange equation for RNA density profile,
we assume $Q(r)$ crosses the value $Q_m$ at most one time.
Certainly, there is a possibility that $Q(r)$ can cross $Q_m$ multiple
times as $r$ increases from zero to $R$. This results in an oscilating
RNA concentration profile. One could easily extend our calculation
presented in this paper to such a case by adding more piecewise solution
to the ansatz, Eq. (\ref{Eq:profile3}), and requiring $Q(r)$
and its derivative to be continuous at the crossing points. 
Such extension could offer insights, for e.g., into the oscillating radial 
profile of RNA molecules packaged inside
Turnip Yellow Mosaic Virus (TYMV)\cite{Jacrot77}. 
Nevertheless, these cases are relatively uncommon and
the calculations would go beyond the scope of this letter. We will
address these cases in more detail in future study.
%
%

Naturally, one wants to know which RNA concentration profile is the most
thermodynamically stable. To answer this question, one needs to
substitute these profiles (Eq. (\ref{Eq:profile2}) and 
Eq. (\ref{Eq:profile3})) into the original expression
for the capsid excess free energy, Eq. (\ref{Eq:HMF}), and compare the
resulting energies. This is a tedious task. Numerically, it is found
that for small adsorption strength of viral capsid,
the second profile would be thermodynamically stable and RNA concentration
is maximum at the capsid center. 
For stronger surface adsorption, the third profile is lower in energy.
In this case, RNA molecules form a dense layer at the capsid
and the RNA concentration is maximum at a finite radius smaller
than $R$.

It is known\cite{NguyenPRL06} that the mean-field theory, Eq. (\ref{Eq:HMF}),
breaks down when the critical point is approached and the first order
transition between dilute and condensed phases of RNA solution becomes
of second order. Once this happens, a physical picture similar
to that of a solution of branched polymer with {\em frozen} branching
arrangement emerges\cite{DaoudJoanny}. In this case,
the RNA molecules become unscreened and non-overlapped. 
For viruses with several packaged RNA molecules, each of them
would adsorb independently onto the capsid and the layer thickness
of each molecule scales as square root of its molecular weight.
Conversely, if such separation between constituent viral genomes is observed,
it would signal the breakdown of mean-field theory.

In conclusion, in this paper we found two different nucleotide concentration
profiles of viral RNA molecules packaged inside spherical viruses. 
The theory applies to a class of viruses where capsid-RNA interaction 
occurs at the capsid surface only. For small interaction strength, the
RNA monomer concentration is maximized at the center of the capsid to maximize
their configurational entropy. For higher interaction strength,
RNA forms a dense layer near the capsid surface. The thickness of
this layer is a slowly varying (logarithmic) 
function of the inner capsid radius. In this case, the amount of
packaged RNA would be proportional to capsid area (or number of capsid proteins)
instead of its volume. The profiles describe reasonably well
the experimental profiles for various viruses.

\begin{acknowledgments}
We would like to thank R. Zhang, V. Belyi, B. Shklovskii, T. Witten
and R. Bruinsma for helpful discussions. 
Initial work on this problem was done with Huaming Li. Nguyen
also acknowledges the Junior Faculty support from the 
Georgia Institute of Technology.
\end{acknowledgments}

\bibliography{RNAsphere}

\end{document}